\documentclass[10pt, letterpaper, conference]{IEEEtran}
    \ifCLASSOPTIONcompsoc
  \usepackage[nocompress]{cite}
\else
  \usepackage{cite}
\fi
\ifCLASSINFOpdf
\else
\fi

\usepackage[english]{babel}
\usepackage{blindtext}
\usepackage{tabularx}
\usepackage[usenames,dvipsnames]{xcolor}
\usepackage{graphicx}
\usepackage{enumitem}
\usepackage{wrapfig}
\usepackage{eso-pic}
\usepackage{graphicx, color, times, amsfonts, amsmath, amssymb}
\usepackage{amssymb}
\usepackage{subfigure}
\usepackage{multirow}
\usepackage{algorithm}
\usepackage{algorithmic}
\usepackage{listings}
\usepackage{array}
\usepackage{epsfig}
\usepackage{soul}
\usepackage{setspace}
\newsavebox{\leftbox}
\newsavebox{\rightbox}
\usepackage{booktabs,siunitx}
\renewcommand{\baselinestretch}{0.96}
\usepackage{booktabs}

\usepackage{pgfplots}
\pgfplotsset{width=7cm,compat=1.8}

\newlist{inlinelist}{enumerate*}{1}
\setlist[inlinelist]{label=(\roman*)}

\definecolor{myblue}{HTML}{5699D6}
\definecolor{myred}{HTML}{F3585C}
\definecolor{mygreen}{HTML}{77C465}
\definecolor{myorange}{HTML}{FBA752}
\definecolor{mypurple}{HTML}{9E64AC}

\begin{document}

\setlength\abovedisplayskip{0.2mm}
\setlength\belowdisplayskip{0.2mm}

\title{Towards a Social Virtual Reality Learning Environment in High Fidelity}

\author{
\IEEEauthorblockN{Chiara Zizza\IEEEauthorrefmark{1}, Adam Starr\IEEEauthorrefmark{2}, Devin Hudson\IEEEauthorrefmark{3}, Sai Shreya Nuguri\IEEEauthorrefmark{4}, Prasad Calyam\IEEEauthorrefmark{4}, and Zhihai He\IEEEauthorrefmark{4}}
\IEEEauthorblockA{
\IEEEauthorrefmark{1}Grinnell College, Email: zizzachi@grinnell.edu\\
\IEEEauthorrefmark{2}Pomona College, Email: acs12014@mymail.pomona.edu\\
\IEEEauthorrefmark{3}Truman State University, Email: dnh3158@truman.edu\\
\IEEEauthorrefmark{4}University of Missouri, Email: snd45@mail.missouri.edu, calyamp@missouri.edu, hezhi@missouri.edu} 
}

\maketitle

\begin{abstract}
Virtual Learning Environments (VLEs) are spaces designed to educate students remotely via online platforms. Although traditional VLEs such as iSocial have shown promise in educating students, they offer limited immersion that diminishes learning effectiveness. This paper outlines a virtual reality learning environment (VRLE) over a high-speed network, which promotes educational effectiveness and efficiency via our creation of flexible content and infrastructure which meet established VLE standards with improved immersion. This paper further describes our implementation of multiple learning modules developed in High Fidelity, a ``social VR'' platform. Our experiment results show that the VR mode of content delivery better stimulates the generalization of lessons to the real world than non-VR lessons and provides improved immersion when compared to an equivalent desktop version.
\end{abstract}

\begin{IEEEkeywords}
Virtual Reality, Virtual Learning Environment, High Fidelity, Multi-user Network Application.
\end{IEEEkeywords}

\IEEEpeerreviewmaketitle

\section{Introduction}  
\label{Sec:Introduction}

\begin{figure}[t]
\centering
\includegraphics[width=\linewidth]{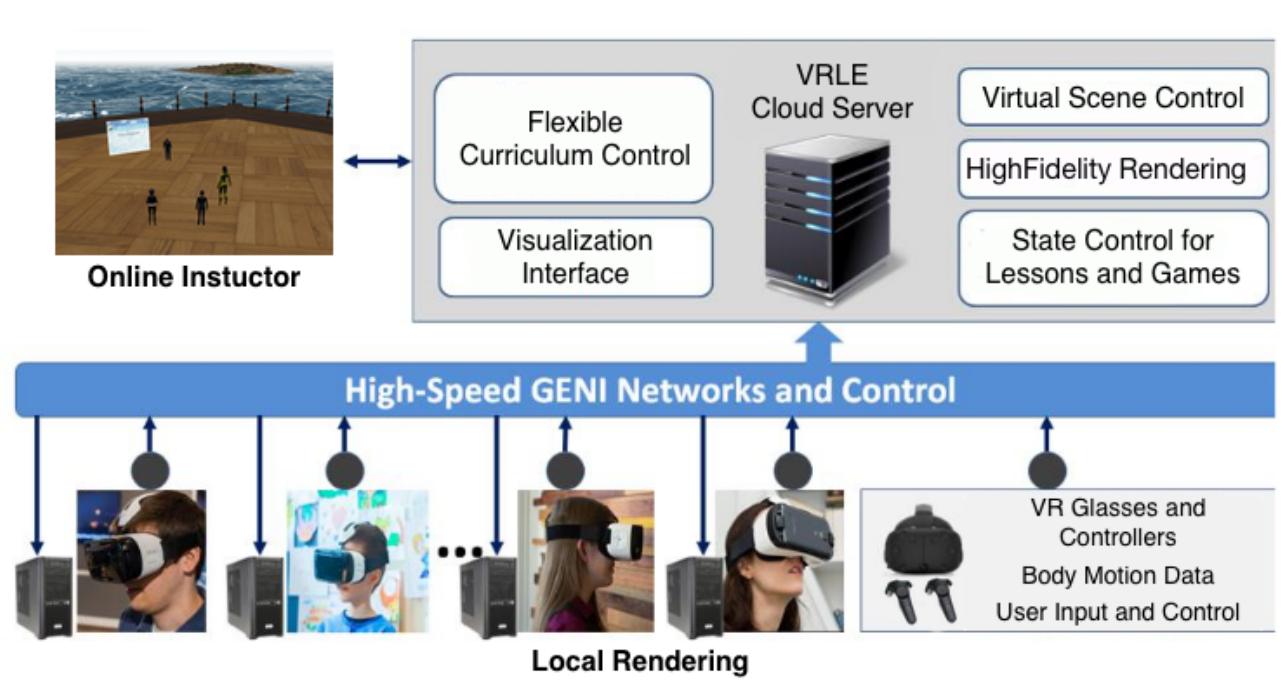}
\caption{{Architecture of a VRLE to implement `social VR' across high-speed networks for online instructor-driven lessons and games using High Fidelity.}}
\label{fig:architecture}
\end{figure}

Past work with Virtual Learning Environments (VLEs) have proven effective in teaching students remotely. One example is iSocial~\cite{isocial}, a desktop application which trains students with Autism Spectrum Disorder to improve their social competencies by enabling social interaction in a virtual world. Virtual reality (VR) has also shown promise in teaching generalizable skills through immersive settings. In this paper, we describe our activities that aim to transform VLEs into ``social VR'' platforms, where multiple, distant users may inhabit the same virtual area and engage in lessons and games around common tasks, as illustrated in Fig. \ref{fig:architecture}. Our prototype platform is a cloud-based and high-speed network-enabled, immersive Virtual Reality Learning Environment (VRLE) where multiple students from across the world may meet with remote instructors.

Our VRLE is built in High Fidelity, an open source client-server software for creating shared VR environments. High Fidelity supports both Desktop and VR modes \cite{HiFi}. The Desktop mode uses the keyboard and mouse, while the VR mode utilizes a headset and two hand controllers and is supported by multiple head-mounted display devices such as the HTC Vive and Oculus Rift. The server runs on a Global Environment for Network Innovations (GENI) cloud rack, which is a distributed computing resource with open access for ``transformative, at-scale experiments in network science, services, and security'' \cite{GENI}. This VRLE will also include real-time analysis of network performance while up to 150 users interact simultaneously within the same High Fidelity environment.

While this project implements iSocial as a VRLE, the structure is generalizable to other contexts with one or more instructor and groups of students. Many VLEs are effective teaching tools, but they often lack a near-realistic learning environment. Our VRLE is designed to provide a more immersive environment, allowing students to more easily connect the lessons and games from the VRLE to real world situations.

The remainder of this paper is organized as follows: Section \ref{Sec:Related_Works} discusses prior works related to VLEs. Section \ref{Sec:problem_statement} lists the problems with VLEs, and describes our solution. Section \ref{Sec:VRLE_Implementation} describes our VRLE implementation. Section \ref{Results and Analysis} details testbed experiment results. Section \ref{Sec:Conclusion} concludes the paper.

\section{Related Works}
\label{Sec:Related_Works}

\cite{schmidt2010social} defines 3D VLEs as ``computer-generated, three-dimensional simulated environments that are coupled with well-defined learning objectives.'' Students enter these environments as avatars via a networked computer and interact with each other with text chat, voice chat, and the manipulation of shared objects. Adding VR to VLEs can offer efficient generalization of skills from a virtual environment to the real world, since the environment mimics real world imagery and context as noted by \cite{vr_for_asd_5}. \cite{vr_for_training_2} presents a review and investigation into VRLE's and the reported benefits these pedagogic systems provide to the students, as well as how they impact students' learning journeys within the program of study. The authors conclude that a correctly implemented VLE improves student performance and is correlated with improved engagement.

\cite{vr_for_training_3} presents a comprehensive review of virtual reality-based instruction research and concludes that VR environments are effective for teaching K-12 and higher education. These results are used by instructional designers to design VRLEs.

Most existing VLEs have been developed as demonstrations, supplements, or assistants for specific training tasks or teacher-led activities, and none of them leverage high-speed networking and cloud computing technologies. In contrast, our VRLE represents an emerging distance education paradigm.

\section{VLE transformation to a VRLE}
\label{Sec:problem_statement}

Research, such as \cite{vr_for_asd_5}, indicates that a near-realistic learning environment is important for students to generalize and transfer the knowledge and skills learned in the virtual space to the real world. However, traditional VLEs lack the capacity for a near-realistic experience. As traditional VLEs (and even some VRLEs such as the one used in \cite{vr_for_asd_5}) utilize a mouse and keyboard, the realism of the environment is limited. For example, iSocial uses highly-structured environments to direct students towards lessons presented as slides or games. While iSocial has some personalization of avatars, it can only represent limited features of the user. iSocial's rigid back-end limits the capacity of instructors to make quick changes to lesson plans. Our VRLE aims to improve the aforementioned limitations and allow for a more effective learning tool.

\subsection{VRLE System Architecture}
\label{VRLE_architecture}
Our VRLE application prototype architecture is consistent with Fig. \ref{fig:architecture} and is designed according to iSocial's standards for virtual learning modules~\cite{isocial}. The standards provide guidance such as: environments built with reduced distractions, realistic avatars, guiding indicators to direct movement, and locking pods to help keep the students in place while viewing a lesson. On the client side, we can support up to 150 users (owing to High Fidelity capabilities and assuming sufficient server-side resources) over different geographical regions, each wearing a VRLE client device. The client device consists of a wearable VR headset, and a local computer. We are adapting the existing iSocial Social Competence Intervention (SCI) training curriculum, as an exemplar learning module, to our new VRLE application. The SCI curriculum is executed in a 3D environment and is globally rendered at the central VRLE cloud server. The rendered content is delivered over high-speed networks to the clients for presentation. The local computers perform on-the-fly content rendering based on the instantaneous user head position and body movement to achieve low-delay, smooth presentation of content on the VR headset. Lesson slides are hosted on a web server and are able to be modified by the instructors themselves. Many of the training lessons are implemented on virtual websites that instructors can control through a consolidated, administrator control panel. The instructor can manipulate course content remotely, allowing students to split into smaller groups for discussion at great enough distances that individual groups cannot hear other conversations.

\subsection{Avatars}
As individualized avatars are a necessary to promote immersion, we utilize Morph 3D's \cite{morph3d} compatibility with High Fidelity to streamline the process of avatar customization and transferring them into the VRLE environment. Morph 3D allows students to quickly customize avatars with many features such as body mass, facial dimensions, and hair color. These avatars can be directly imported into High Fidelity, which gives students a sense of individuality without requiring knowledge of 3D rendering software. Morph 3D allows easy avatar creation with a greater capacity for personalization than is available in iSocial and other VLEs.

\subsection{Web Apps}
Additionally, we have created a web platform that hosts instructional web pages. Browsers located in teaching areas create convenient and intuitive classrooms. The platform allows for lessons to be easily added and changed by instructors more efficiently than iSocial. The browsers access different web pages depending on the intended activity. For example, if an instructor wants to present a slide-show, they may select the slide viewer web site from the homepage. Whereas, if the instructor wants to play a specific game to test the students' recently acquired knowledge, they can navigate to that site.

The websites are configured to allow only administrators to change the state. This restriction is necessary in contexts where students may be disruptive or may accidentally press buttons in the environment. 

\begin{figure}[htb!]
\centering
\includegraphics[width=\linewidth]{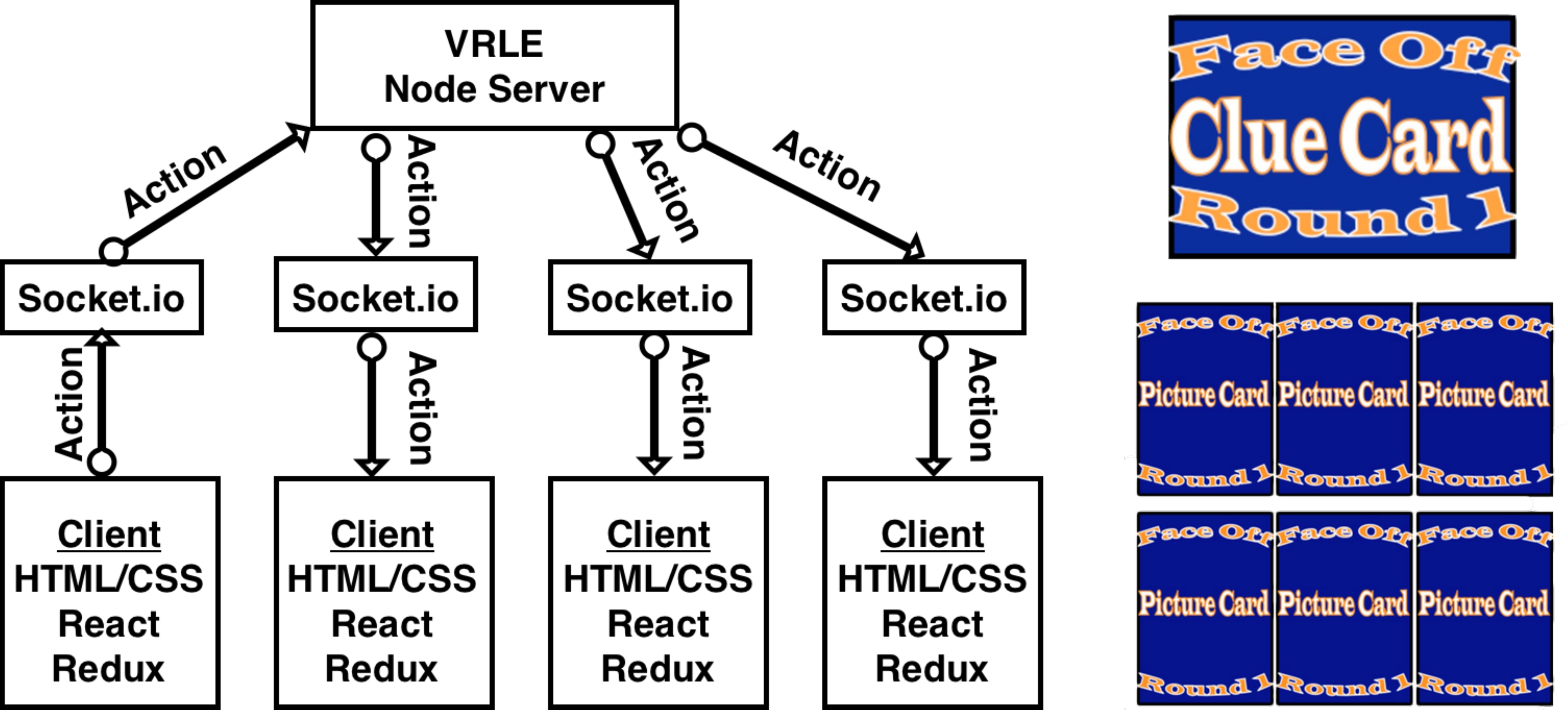}
\caption{{Left: Architecture of the web pages for instruction content. This diagram shows the path of a state change action. The action is sent from one client and broadcast to all other clients. Right: example web page of the Face Off game.}}
\label{fig:webDesign}
\end{figure} 

The web pages for the instructional content are built with the React framework, and the state is maintained across browser instances by the VRLE node server with a combination of Redux and Socket.io as shown in Fig. \ref{fig:webDesign}. An action in one client is then sent to the VRLE node server and broadcast to all other clients. Clients only update their states if the action is relevant to their current display. Slide-show presentations are managed by a React-Router program to easily select presentations to display in in the VLE. 

As every user in High Fidelity sees a different instance of a web page, all instances must be synced using Socket.io. This implementation via React, Redux, and Socket.io is general enough to be easily adapted for many contexts. By replacing the images or adjusting a few lines of code, the same program may work in many different contexts allowing the reuse of a substantial amount of code. In this way, the VRLE permits the easy inclusion of additional activities and lessons. 

The administrator controls are implemented as a virtual 'tablet application' in High Fidelity. In High Fidelity, users may pull up a virtual tablet to change settings or use installed applications, such as the VRLE administrator controls. This application is only available to the instructors and connects to the VRLE server through Socket.io. The application sends Redux actions to the various games and sideshows. This tablet approach limits the potential for accidental or malicious state changes.   

Not only is this web framework convenient to display content, it is necessary to reference common web pages as High Fidelity renders a unique site instance for each user in the environment. Our web app framework syncs all instances to maintain a common state. 

\subsection{Virtual Reality Environment}
\begin{figure}
\centering\includegraphics[width=\linewidth]{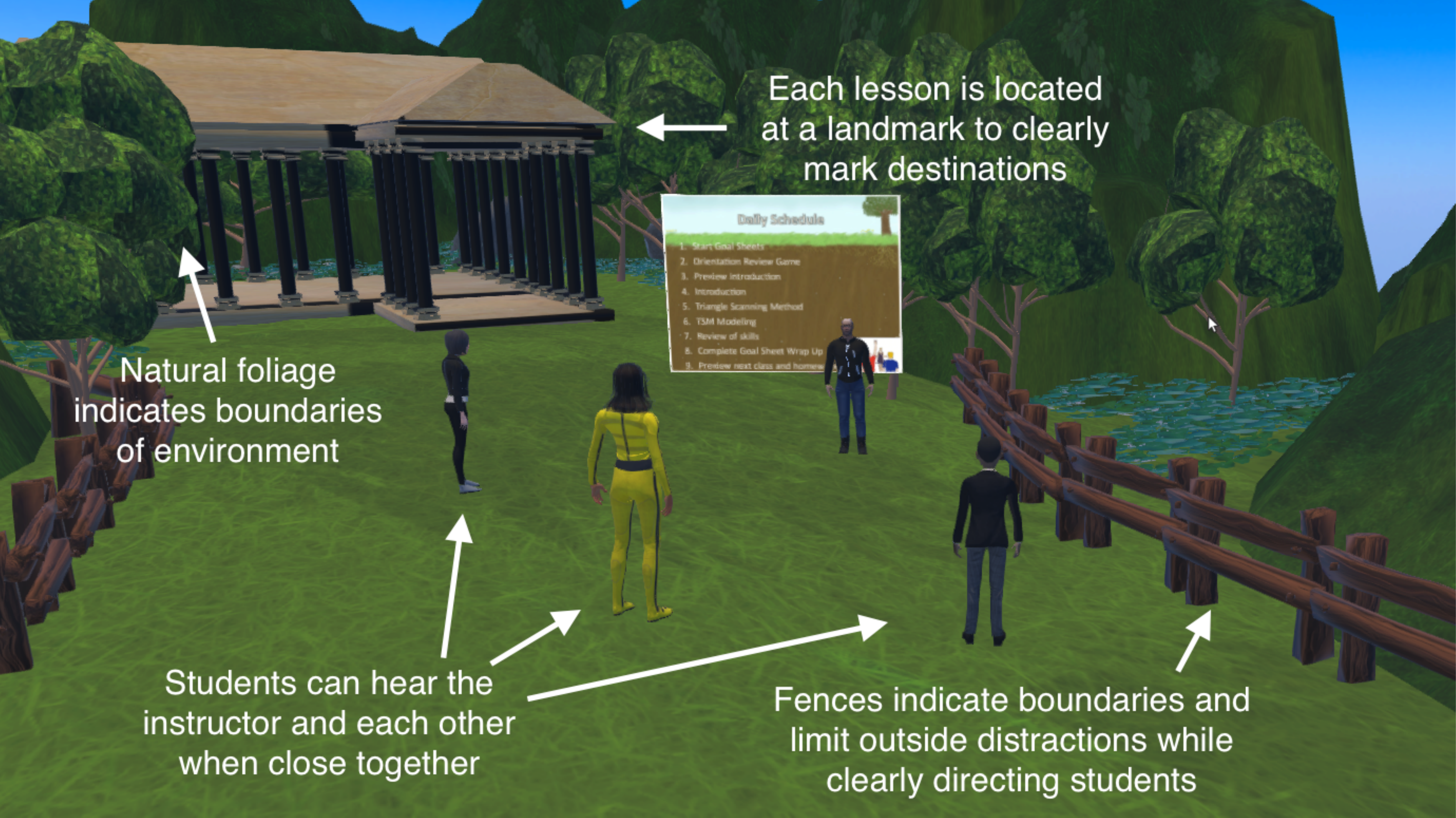}
\caption{{Overlook of a iSocial standards compliant VRLE module demonstration with slide show and avatars of an instructor and three students.}}
\label{fig:Unit2}
\end{figure}
Fig. \ref{fig:Unit2} provides an outlook of one lesson of our VRLE, which is compliant with iSocial standards. The fundamental idea for the VRLE's layouts is to have open areas in which the students' freedom to move and interact is maintained, with natural foliage and rocks acting as clear boundaries. Each lesson of our VRLE has a unique, designated area with the theme typically reflecting the topic being taught. For example, a lesson might have the students determine individual roles on a deserted island (e.g. water collector, lookout) to practice cooperation and teamwork. Each lesson also has a central area with an interactive web page, which the instructor controls, as a way of directing the students towards a common focus. Clear pathways, outlined with fences, connect different lesson parts. Teleportation portals connect related lessons and provide a quick, convenient way of traveling distances in the VRLE. Each lesson also has a unique name to facilitate direct access via teleportation.

A key component to our VRLE is that the environment is tangible and reacts to the student. When applicable, students are able to interact with various objects in the world and manipulate certain components. 
High Fidelity also allows users in an environment to speak with one another. The adjustable attenuation coefficients allow control of how audio levels diminish over distance, which gives the realistic effect of the speaker sounding quieter, the farther they are from the listener. This feature aids in keeping the students by the instructor and allows small group activities where students may discuss privatively before returning for a full-group debrief.

\subsection{Orientation for Virtual Reality}
As students in our VRLE may not have experience in VR, the first lesson is an orientation. In the first part of the VRLE's orientation program, students create personalized avatars for themselves. The second part teaches the students how to operate the VR controllers and navigate the VRLE in a fun, hands-on way. Once done with the tutorial, the students move onto the lesson modules.

\section{VRLE Implementation}
\label{Sec:VRLE_Implementation}

\begin{figure}[htb!]
\centering
\includegraphics[width=\linewidth]{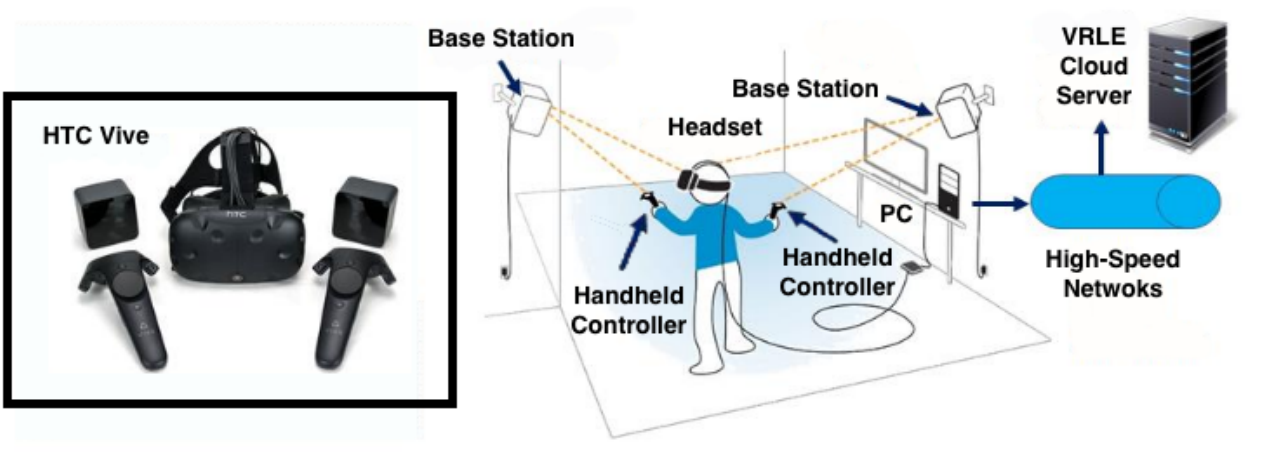}
\caption{{Major components of the experimental testbed setup for the usability study with HTC Vive and VRLE cloud server.}}
\label{fig:components}
\end{figure}

Fig. \ref{fig:components} shows the major components of our proposed VRLE client device and the set up. For the immersible VR access, we are currently using the latest HTC Vive system \cite{HTC2016}, which has a VR headset, two hand-held controllers, and two base stations for accurate localization and tracking of the controllers. The headset has a wired connection to a computer, which communicates with the VRLE cloud server on GENI through high-speed networks. The headset tracks head pose, displays interactive and immersive course content, and responds to head motion in real-time. Users interact with the environment, instructors, and peers using their virtual hands (controllers) and voices. The HTC Vive provides adequate tracking accuracy, flexibility, resolution, system stability, cost-effectiveness, and user comfort for our field experimentation. 

\vspace{-3mm}
\vspace{-3mm}

\section{Results and Analysis}
\label{Results and Analysis}

Qualitative tests were conducted with an expert and architect of iSocial to obtain preliminary evaluations of our VRLE application prototype. The expert was shown the VRLE with one remote instructor and two students equally distant from the instructor connected over a high-speed network. The expert concluded that: the VRLE is more immersive than iSocial, layout is more engaging while not distracting, and the web app architecture is reusable for many of the original iSocial lessons.

\begin{figure}[htb!]
\centering
\begin{tikzpicture}[scale=.95]
\begin{axis}[
	title  = Responses to ``I feel present in \lbrack the VRLE\rbrack'',
    title style={align=center},
    ybar stacked,
    x=0.105\textwidth,
	bar width=0.07\textwidth, 
    enlargelimits=0.20,
    legend style={at={(1.05,.5)},
      anchor=west, font=\fontsize{8}{5}\selectfont},
    ylabel={Proportion of Subjects},
    symbolic x coords={Desktop, VR},
    xtick pos=left,
	ytick pos=left,
    xtick=data,
    width=.2\textwidth,
    height=.3\textwidth,
    enlarge x limits=0.5,
	nodes near coords, 
    nodes near coords align={anchor=north},
    totals/.style={nodes near coords align={anchor=south}},
    nodes near coords/.style={color=black},
    every node near coord/.style={color=black},
    reverse legend,
    ]
\addplot+[ybar, fill=myblue] plot coordinates {(Desktop,0) (VR,0) };
\addplot+[ybar, fill=myorange, point meta=explicit] plot coordinates {(Desktop,0.2857142857)[0.29] (VR,0)[0] };
\addplot+[ybar, fill=myred, point meta=explicit] plot coordinates {(Desktop,0.4285714286)[0.43] (VR,0)[0] };
\addplot+[ybar, fill=mygreen, point meta=explicit] plot coordinates {(Desktop,0.2857142857)[0.29] (VR,1)[1] };
\addplot+[ybar, fill=mypurple, point meta=explicit] plot coordinates {(Desktop,0)[0.29] (VR,0)[1] };
\legend{\strut Strongly Disagree, \strut Disagree, \strut Neither Agree nor Disagree, \strut Agree, \strut Strongly Agree}
\end{axis}
\end{tikzpicture}
\caption{{Responses to ``I feel present in [the VRLE]'' for Desktop and VR tests indicate unanimous sense of presence in VR i.e., we observed increased \textit{engagement} in the VRLE module from Desktop to VR.}}
\label{fig:present}
\end{figure}
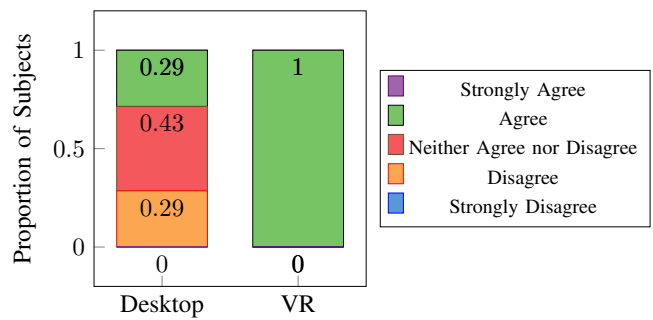

\begin{figure}[htb!]
{
\begin{tikzpicture}[scale=.91]
\begin{axis}[
	title  = Responses to ``I enjoyed my time in \lbrack the VRLE\rbrack'',
 	title style={align=center},
    ybar stacked,
    x=0.105\textwidth,
	bar width=0.07\textwidth, 
    enlargelimits=0.20,
    legend style={at={(1.05,.5)},
      anchor=west, font=\fontsize{8}{5}\selectfont},
    ylabel={Proportion of Subjects},
    symbolic x coords={Desktop, VR},
    xtick pos=left,
	ytick pos=left,
    xtick=data,
    width=.2\textwidth,
    height=.3\textwidth,
    enlarge x limits=0.5,
	nodes near coords, 
    nodes near coords align={anchor=north},
    totals/.style={nodes near coords align={anchor=south}},
    nodes near coords/.style={color=black},
    every node near coord/.style={color=black},
    reverse legend,
    ]
\addplot+[ybar, fill=myblue] coordinates {(Desktop,0) (VR,0) };
\addplot+[ybar, fill=myorange] coordinates {(Desktop,0) (VR,0) };
\addplot+[ybar, fill=myred, point meta=explicit] coordinates {(Desktop,0.1428571429) [0.14] (VR,0) [0] };
\addplot+[ybar, fill=mygreen, point meta=explicit] coordinates {(Desktop,0.8571428571)[0.86] (VR,0.1428571429)[0.14] };
\addplot+[ybar, fill=mypurple, point meta=explicit] coordinates {(Desktop,0)[0.86] (VR,0.8571428571)[0.86] };
\legend{\strut Strongly Disagree, \strut Disagree, \strut Neither Agree nor Disagree, \strut Agree, \strut Strongly Agree}
\end{axis}
\end{tikzpicture}
\caption{Responses to ``I enjoyed my time in [the VRLE]'' for Desktop and VR tests show strong \textit{enjoyment} in both Desktop and VR with a unanimous increase in enjoyment in VR i.e., we verified \textit{minimal distraction} in the VRLE module with greater \textit{immersion} in VR.}
}
\label{fig:enjoy}
\end{figure}
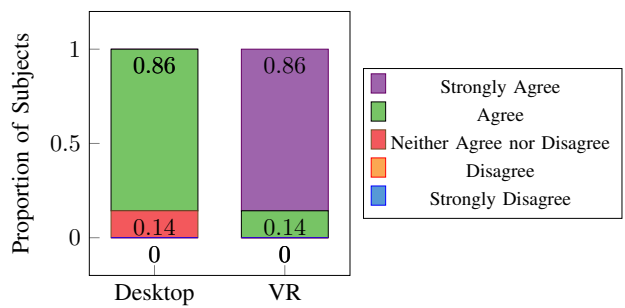

Subjective tests for quantitative assessment were conducted with seven typically developing college-age subjects, comparing mouse and keyboard movement with VR hand-held controller movement. Subjects were randomly assigned a condition of testing desktop or VR first. Each subject was led around the VRLE and engaged in activities such as a slide show and the Face Off game. A survey was taken after each test (twice per subject). 

When operating on the desktop, subjects expressed frustration with mouse and keyboard movement, as they found movement slow and tedious. When transitioning to VR in our VRLE, the subjects found the teleportation mechanism preferable. Additionally, subjects indicated that VR movement was more immersive and fluid: head movement directed the avatar's orientation and hand movements were reflected on the avatar as one typically expected. Responses indicate that the subjects feel more present in VR. In fact, every subject indicated that they felt present in the VRLE in the VR mode (see Fig. \ref{fig:present}), compared to engagement in the desktop being less than half that amount. Every subject enjoyed VR more than desktop as show in Fig. \ref{fig:enjoy}, indicating the greater immersion of the VRLE, and when asked, subjects unanimously stated that they preferred VR. The VR mode is not perfect; for example, about a third of the subjects felt dizzy in VR at times. Likely, this sensation was due to the subject moving their avatar with thumb-pad movement as the discomfort was corrected with use of teleportation.

\section{Conclusion and Future Work}
\label{Sec:Conclusion}
In this paper, we overcome shortcomings in traditional VLEs with an early prototype implementation of an immersive VRLE in High Fidelity with the following capabilities: users can talk with each other, split into groups to discuss privately as teams, and share educational resources such as slide shows and games. The instructor can control course content from a central administrator application. Preliminary qualitative tests with an iSocial VLE expert showed that our VRLE is more immersive than iSocial VLE, while not being distracting. The expert concurred that that our web architecture allows many activities to be implemented with relative ease. Preliminary subjective tests confirm that the VRLE is more immersive and enjoyable compared to VLEs controlled by mouse and keyboard. Users also felt `present' in the VRLE and experienced minimal distractions, owing to our compliance with established VLE lesson design standards in iSocial. 

To the best of our knowledge, our work is the first to detail a VRLE architecture in a social VR paradigm that is supported by platforms such as High Fidelity. Our work also builds a foundation upon which our VRLE lessons can now be enhanced for social VR and used in realistic lesson delivery with actual users who are geographically distributed. Our upcoming plans to extend this work include: \begin{inlinelist}
\item optimizing network performance,
\item analyzing student engagement via electroencephalogram (EEG) headbands,
\item tracking progress across lessons and performance across sessions on a custom social network web App,
\item utilizing the capabilities of the Microsoft Kinect \cite{kinect} to track body movements, and
\item using face-tracking VR headsets, such as Veeso \cite{veeso}, to track facial expressions to help instructor assessment of student learning.
\end{inlinelist}
\section*{Acknowledgment} 
\addcontentsline{toc}{section}{Acknowledgment}
This material is based upon work supported by the National Science Foundation under Award Numbers: CNS-1647213 and CNS-1659134. Any opinions, findings, and conclusions or recommendations expressed in this publication are those of the authors and do not necessarily reflect the views of the National Science Foundation.

\bibliographystyle{IEEEtran}
\bibliography{IEEEabrv,VRLE}

\end{document}